\documentclass{article}
\usepackage[T1]{fontenc}
\usepackage[utf8]{inputenc}
\usepackage{ismir,amsmath,cite,url}
\usepackage{graphicx}
\usepackage{color}
\usepackage{subcaption}
\usepackage{amssymb}

\usepackage{mathastext}

\usepackage{booktabs}

\usepackage{array}
\newcommand\Tstrut{\rule{0pt}{2.2ex}}

\usepackage{lineno}
\nolinenumbers

\title{Adapting Meter Tracking Models to Latin American Music}

\multauthor
{Lucas S. Maia$^1$ \hspace{.5cm} Martín Rocamora$^2$ \hspace{.5cm} Luiz W. P. Biscainho$^1$ \hspace{.5cm} Magdalena Fuentes$^3$}{
$^1$ PEE/COPPE, Federal University of Rio de Janeiro, Brazil\\
$^2$ FING, Universidad de la República, Uruguay \\
$^3$ MARL-IDM, New York University, United States \\
{\tt\small lucas.maia@smt.ufrj.br}
}

\def\authorname{L.S. Maia, M. Rocamora, L.W.P. Biscainho, and M. Fuentes}

\usepackage[bookmarks=false,pdfauthor={\authorname},pdfsubject={\papersubject},hidelinks]{hyperref}
\begin{document}

\maketitle

\begin{abstract}

Beat and downbeat tracking models have improved significantly in recent years with the introduction of deep learning methods. However, despite these improvements, several challenges remain. Particularly, the adaptation of available models to underrepresented music traditions in MIR is usually synonymous with collecting and annotating large amounts of data, which is impractical and time-consuming. Transfer learning, data augmentation, and fine-tuning techniques have been used quite successfully in related tasks and are known to alleviate this bottleneck. Furthermore, when studying these music traditions, models are not required to generalize to multiple mainstream music genres but to perform well in more constrained, homogeneous conditions. In this work, we investigate simple yet effective strategies to adapt beat and downbeat tracking models to two different Latin American music traditions and analyze the feasibility of these adaptations in real-world applications concerning the data and computational requirements. Contrary to common belief, our findings show it is possible to achieve good performance by spending just a few minutes annotating a portion of the data and training a model in a standard CPU machine, with the precise amount of resources needed depending on the task and the complexity of the dataset.

\end{abstract}

\section{Introduction}\label{sec:introduction}

Meter tracking means following the pulsating temporal structure of music from audio signals, which implies identifying at least beats and downbeats~\cite{Klapuri2006meter}. It is a long-standing area of research in music information retrieval (MIR) with applications ranging from automatic DJ mixing~\cite{Chen2022automaticdj} to musicological studies~\cite{Srinivasamurthy2017frontiers}. Meter tracking has gone through a big transformation in the last decade due to the introduction of deep learning (DL) techniques~\cite{Bock2016, Durand2017, fuentes2018, heydari2021}, which brought an improvement in performance as well as a change in the design paradigm of related methods~\cite{tutorialISMIR2021}. Nowadays, beat and downbeat tracking models rely mostly on supervised DL~\cite{tutorialISMIR2021}, and thus become data-driven and requiring large amounts of annotated data to generalize to different songs, genres or datasets.

This dependence on annotated data poses many challenges to the widespread use and adoption of such models, especially for culturally specific music traditions~\cite{Silla2008, cano2020, sarria2019}, which often lack annotated data as producing annotations requires culturally-aware expertise. For this reason, off-the-shelf general-purpose models typically underperform in these music genres since they are underrepresented in the datasets used for training. Nevertheless, previous work on some Latin American music traditions shows that if annotations are available, training statistical models % (i.e., not deep learning-based) 
can produce good performance results~\cite{Nunes2015candombe}. 

Recent works have started to look at beat tracking from a different perspective. Instead of developing ``universal'' models capable of performing equally well across various music genres (requiring large quantities of labeled data), recent efforts have shifted towards adapting preexisting models to succeed on a subset of interest~\cite{Fiocchi2018finetuning}, which can be as restricted as a single musical piece~\cite{Pinto2021finetuning, yamamoto2021}. This paradigm aligns well with real-world applications, where it is reasonable for a user to spend a short time producing a few seconds of annotations to get a good performance. 

We apply this idea to the refinement of a meter tracking model so that it works well in a particular music genre. We argue that if the genre presents enough homogeneity in terms of its instrumentation and metric structure, as is the case with many Latin American music traditions, it is possible to adapt meter tracking models to perform notably well with just a few annotated data points. We explore this adaptability in terms of data, performance and computational cost. While focusing on two Latin American music genres, \emph{samba} and \emph{candombe}, we study the adaptation of a deep learning state-of-the-art model~\cite{Bock2020tcn} and compare it with a simpler statistical model~\cite{Krebs2013bayesbeat}. Our contributions are: 
1) We perform a detailed analysis on how much annotated data and computation time in CPU are needed to achieve close to ``full-dataset'' performance in \emph{samba} and in \emph{candombe}, including models trained from scratch and fine-tuned, and compare them to off-the-shelf models trained with Western music; 2) We propose initial experiments to understand the homogeneity conditions under which this adaptation will be successful; 3) We open-source our experiments and provide pre-trained models.

\subsection{Other adaptive methods}

Existing adaptive methods typically feature some form of transfer learning or fine-tuning, use deep learning models, and are concerned with either beat tracking~\cite{Pinto2021finetuning, Fiocchi2018finetuning, yamamoto2021} or onset detection~\cite{Fonseca2021finetuning}. Fiocchi et al.~\cite{Fiocchi2018finetuning} adapt a beat tracking model via transfer learning from Western music to a dataset of Greek music. The authors explore both recurrent neural networks (RNNs) and long-short term memory networks (LSTMs), plus a dynamic Bayesian network (DBN) for inference. The model is fine-tuned using a big training set, though with limited success which the authors attribute to the challenges of the dataset. Even though this is an interesting approach for exploring the idea of adapting to a particular music genre, RNNs and LSTMs are known to be computationally expensive, so in the context of real-world model adaptation they are a concerning choice. 

On the other hand, Pinto et al.~\cite{Pinto2021finetuning} and Yamamoto~\cite{yamamoto2021} explore temporal convolutional networks (TCNs), and focus on the adaptation of models to a particular piece of interest of the user. These authors showed that is possible to adapt TCN models using very small quantities of data (in the order of seconds) to work well, in particular with musically challenging pieces. Furthermore, the TCN is a light-weighted model, computationally more efficient. 

Finally, Fonseca et al.~\cite{Fonseca2021finetuning} apply similar ideas to the adaptation of an onset detection model (also featuring TCNs) to a Latin American music tradition: \textit{maracatu de baque solto}. The authors fine-tune the last layers of the TCN with just a few seconds of manual annotations, and show the advantage of instrument-specific models for the automatic annotation of onsets for musicological studies.

\subsection{Latin American Music Traditions}
\label{subsec:afro_music}

\emph{Candombe} drumming is a musical tradition from Uruguay that constitutes an essential part of its popular culture and African heritage. Its rhythm is structured in 4/4 meter, and it is played while marching in the streets using three types of drums of different sizes and pitches: \emph{chico}, \emph{repique}, and \emph{piano}. Each of these drums has a distinctive rhythmic pattern and musical role. An additional time--line pattern, called \emph{clave} or \emph{madera}, is shared by the three drums. The \emph{chico} drum is the timekeeper; it repeats a one-beat pattern that establishes the pulse throughout the performance. The \emph{repique} drum is the improviser; it alternates \emph{clave} patterns and characteristically syncopated phrases. The \emph{piano} drum delineates the timeline with distinctive one-cycle patterns and occasionally interposes ornamented \emph{repique}-like figurations. The rhythm shares many traits with other musical traditions of the Afro-Atlantic world. Notably, some of its rhythmic patterns have strong phenomenological accents displaced with respect to the metric structure and divide the rhythmic cycle irregularly with few strokes on the beat. 

 In parallel, \emph{samba} is a Brazilian musical genre deeply rooted in Brazilian culture, and also has African origins. The word ``\emph{samba}'' actually describes a family of different subgenres, the most famous arguably being \emph{samba de enredo}, \emph{partido alto}, \emph{bossa nova}, and \emph{pagode}. Similarly to \emph{candombe}, \emph{samba} can be played while parading, which is most common during the festivities of \emph{Carnaval}. It can also be performed in more informal settings, in \emph{rodas} and bars, or even as a chamber-music-like style. Its rhythm is commonly perceived in 2/4 meter, and is conveyed by several types of percussion instruments --— \emph{tamborim}, \emph{pandeiro}, \emph{surdo}, \emph{cu\'ica}, \emph{agog\^o}, among others. Each instrument has a handful of distinct patterns~\cite{Goncalves2000}, and more than one instrument may act as the timekeeper. Because of this combination of timbres and pitches, the texture of a performance can become very complex. \emph{Samba} has unmistakable characteristics as the strong accent on the second beat and the development of contrametric structures. 

\section{Method}
Following our intuition about the high homogeneity of \emph{candombe} and \emph{samba} as discussed in Section \ref{subsec:afro_music}, our objective is to understand if it is possible to train meter tracking models with small quantities of data from these music traditions, and if so, how much is needed. To that end, we train the models with increasing amounts of annotated data, ranging from less than a minute up to nearly 40~min, %\qty{40}{\minute}, %the full dataset,
and compare the performance and computational cost of each configuration against the others. We contrast three different training strategies: 1) training the model from scratch with either \emph{candombe} or \emph{samba} snippets; 2) fine-tuning a model trained with 38~h %\qty{38}{\hour} 
of data from diverse datasets of Western music to work on either \emph{candombe} or \emph{samba}; and 3) same as the previous two, %as above,
but training the models with data augmentation to artificially increase the ``small data'' input. %  given to the models. 
 We use a state-of-the-art temporal convolutional network model~\cite{Bock2020tcn} for our experiments, as it presents a good compromise between performance and computational cost. We contrast this model against off-the-shelf models trained in Western music. To understand the adaptability and computational cost of deep learning based methods, we compare the TCN against another simple yet effective baseline, a Bayesian model (BayesBeat)~\cite{Krebs2013bayesbeat}. In the following, we explain our methodology in detail.
 
 \subsection{Datasets}
\label{ssec:datasets}

We have selected datasets of two different Afro-rooted Latin American music traditions for our experiments. First, the \emph{Candombe} dataset~\cite{Nunes2015candombe, Rocamora2015audio}, which consists of 35 recordings of \emph{candombe} drumming, for a total of nearly 2.5~h. % \qty{2.5}{\hour}.
Each track contains an ensemble recording of three to five drummers using different configurations of drums. Tempo varies greatly and often increases along the performance. To represent the \emph{samba} genre, we use the ``acoustic mixtures'' data from the BRID dataset~\cite{Maia2018brid}. These correspond to 93 %\num{93}
short tracks (about 30~s %\qty{30}{\second} 
each) of musicians playing together rhythm patterns found in \emph{samba} and two of its subgenres (\emph{samba de enredo} and \emph{partido alto}). Ten different instrument classes are represented, and two to four musicians take part in each track. The dataset contains a variety of tempi; however, tempo remains fairly constant within each track. In order to consistently train our models with about the same amount of data from both \textit{candombe} and \textit{samba}, and also to allow the comparison between the results obtained for both sets, \emph{candombe} tracks were segmented into non-overlapping 30~s %\qty{30}{\second} 
excerpts. In each experiment repetition, we use a sample of 93 %\num{93} 
\emph{candombe} excerpts.

We also used six datasets to train the baseline TCN model: Ballroom~\cite{Gouyon2006ballroom, Krebs2013bayesbeat}, Beatles~\cite{Davies2009metrics}, GTZAN~\cite{Tzanetakis2002gtzan, Marchand2015swing}, and RWC (Classical, Popular, Jazz)~\cite{Goto2002rwc,Goto2004rwc}. These are commonly used in meter tracking tasks and together correspond to over 38 h of audio data. The Ballroom and GTZAN datasets comprise many diverse music genres (e.g., waltz, tango, rumba, rock, pop, country, etc.). We used the loaders from \textit{mirdata v0.3.6} \cite{mirdata}, except for a custom loader used with Ballroom.

\subsection{Working with small size datasets}
\label{ssec:small_data}

For our experiments, in all cases, we first separate train and test data (80\% and 20\% of 93 %\num{93}
excerpts respectively) to ensure a fair assessment of the models. Then, we divide the training data into six subsets, spanning \{4, 9, 18, 37, 55, 74\} 30-second tracks. We want to determine how differently the models adapt to small quantities of data, so we followed a similar approach to that of~\cite{Pinto2021finetuning} to define the amount of data to be used for training. We select %ed
short 10~s %\qty{10}{\second}
temporal regions at the beginning of the audio excerpts, along with the corresponding beat and downbeat annotations, and discard the remaining audio portion. Then we split each of these regions into two adjacent 5~s %\qty{5}{\second}
parts, the first to be used for training and the second reserved for validation in the TCN model; alternatively, we use the entire 10~s %\qty{10}{\second}
for training the Bayesian % whenever the
model % worked
with off-the-shelf parameters. Considering that each snippet only lasts {10~s}, %\qty{10}{\second}, 
these data subsets add up to approximately 40~s, %\qty{40}{\second},
 1.5, 3, 6, 9, and 12~min % \qty{12}{\minute}
 of annotations, respectively. %(see a summary in Table~\ref{tab:strategies}).
The rationale behind this strategy is that given a set of recordings of such Latin American music traditions in real-world applications, it would be reasonable to ask a user to annotate just a few seconds to a few minutes of data; of course, the less data needed, the better.

Given that we are using very few data points to train the models, performance is strongly affected by data sampling. To mitigate this, we repeat all of our experiments 10 times with different seeds for the random data split generation, which means that models are trained 10 times with each of the different subset sizes. Note that selecting the best strategies for data sampling is out of the scope of this work, and left to be addressed in the future. Test data are left uncut, i.e., we use the full 30~s, %\qty{30}{\second},
to keep compatibility with common model evaluation practices in meter tracking. % in the evaluation of the models, we use the full \qty{30}{\second} to keep compatibility with common practices in meter tracking.

\subsection{TCN Model}
\label{ssec:tcn}

We use in our experiments the TCN multi-task model presented in \cite{Bock2020tcn}, in particular the open-source implementation of \cite{tutorialISMIR2021}. % The TCN model architecture consists of 11 dilated convolution layers with increasing dilation rates, which enable exponentially large receptive fields and thus can incorporate both large and short temporal contexts. This multi-scale temporal capacity of the TCN is key for reliably estimating both beats and downbeats. The input of this model is a log-magnitude spectrogram with 81 logarithmically spaced frequency bins and a frame rate of 100 frames per second. As per the multi-task formulation, it has been shown in \cite{Bock2020tcn} that it improves the individual estimation of beats, downbeats, and tempo.
In this work, we focus on meter tracking, and ignore the tempo estimation head of the network. First, the TCN estimates the beat and downbeat likelihood. Then, we use two different implementations of a DBN (\textit{DBNBeatTracker} and \textit{DBNDownBeatTracker} from \textit{madmom v0.17.dev0}~\cite{madmom}) to infer the final positions of beats and downbeats respectively. Inferring them separately rather than jointly led to better results.

\subsection{Training strategies}

\subsubsection{Training from scratch (TCN-FS)}
\label{sec:from_scratch}

For datasets with high similarity in terms of instrumentation, rhythmic patterns, and tempo, we expect that we can train a model from scratch with a few training points that would work well for most of the data. 

Following the explanation in Section \ref{ssec:small_data}, we train one model per data subset, % depicted in Table \ref{tab:strategies},
and repeat this 10 times with randomly initialized weights and seeds. We also consider the case in which all annotations are available and include the analysis of model performance when training with the entire 30-second excerpts. In this situation, we split the 74 %\num{74}
train excerpts into train and validation (75\%/25\%). For every strategy, % W % added
we use a learning rate of 0.005, and reduce it by a factor of 0.2 if validation loss did not improve after 10 epochs. We train for a maximum of 100 epochs, early stopping at 20 epochs.

% \textcolor{red}{learnrate, epochs, lrfactor, lrpatience, espatience = 0.005, 100, 0.2, 10, 20}

%\begin{table}[!ht]\small
%    \centering
%    \begin{tabular}{c}
%       \hline
%        Annotation (min) \Tstrut \\
%       \hline
%       \{0.67, 1.5,   \Tstrut%\\
%       3.0, 6.17, 9.17 %\\
%       12.33, all\}   \\
%        \hline
%    \end{tabular}%\hspace*{15pt}
%    \begin{tabular}{c}
%       \hline
%         Training \Tstrut \\
%       \hline
%        from scratch (FS) \Tstrut\\
%        fine-tuning (FT) \\
%        data augmen.~(A) \\
%        \hline
%    \end{tabular}
%    \caption{Annotated minutes in small-data subsets.}% Small-data subsets and training strategies.}
%    \label{tab:strategies}
%\end{table}

\subsubsection{Fine-tuning (TCN-FT)}
\label{sec:fine_tuning}

We also approach the problem of meter tracking in a culture-specific setting from a ``transfer learning'' perspective. Following~\cite{Fiocchi2018finetuning, Fonseca2021finetuning, Pinto2021finetuning}, we adapt a meter tracking model that was previously trained for a different musical context. The intuition here is that if the model is first trained on a large dataset, even if it was built around Western music, it can serve as a good starting point for a model that is to be tuned for a specific out-of-training music tradition. 
This is a realistic approach since most of the available annotated data and trained models are Western-based. % added after R2's comments
For this purpose, we trained a baseline TCN model on the Ballroom, Beatles, GTZAN, and RWC datasets. Due to the nature of its training data, this baseline model has to cope with many different meters, genres and acoustic conditions, which makes it a good starting point. We fine-tuned it by using the same training procedure described previously with the initial learning rate reduced to 0.001, a fifth of the value used in the FS approach, as in \cite{Pinto2021finetuning}. 

\begin{figure*}[ht]
    \centering
    \includegraphics[scale=.51]{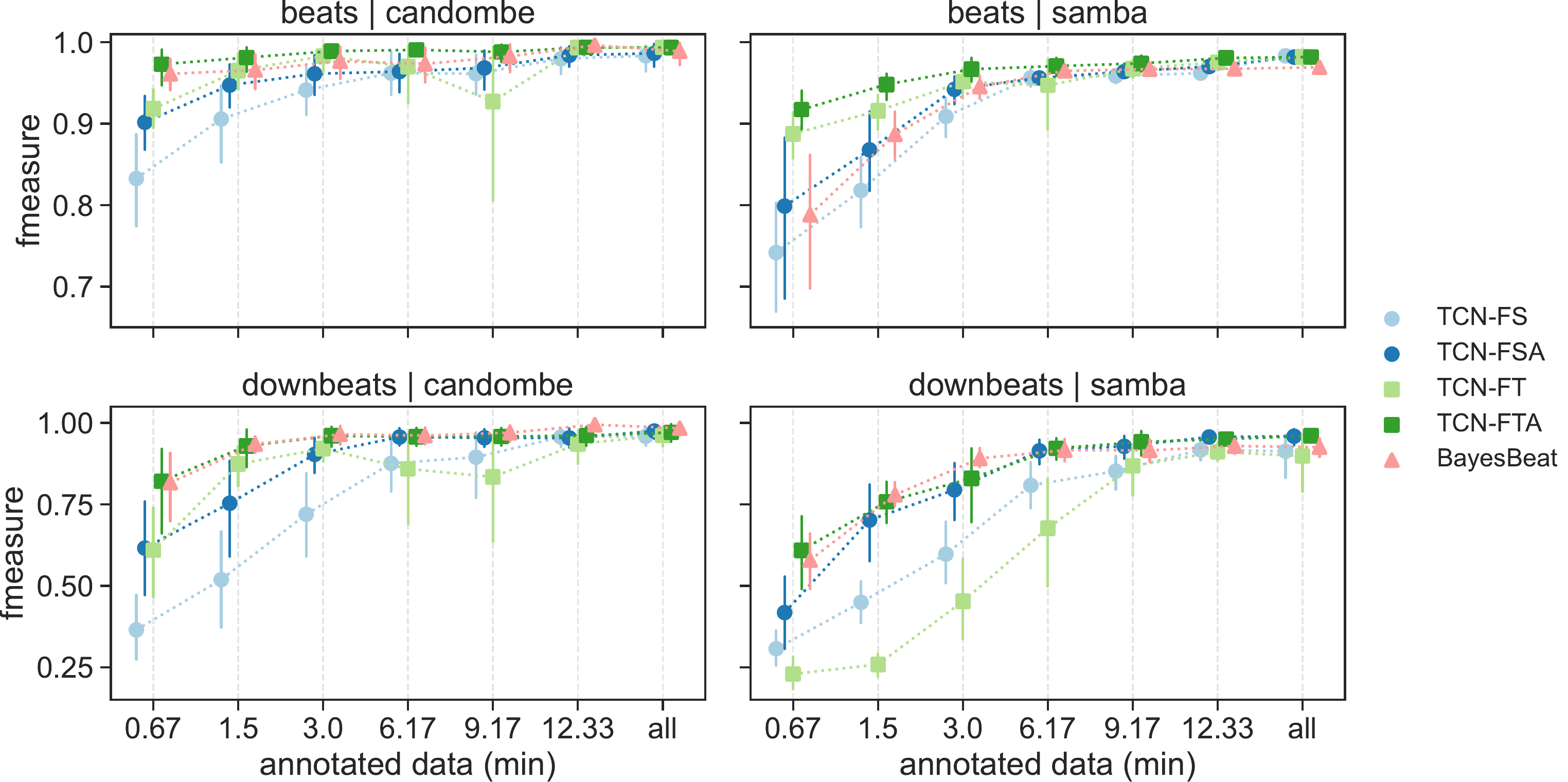}
    \caption{Performance of different model and training configurations. Label ``all'' indicates fully-annotated dataset.}
    \label{fig:fmeasure}\vspace*{-.5\baselineskip}
\end{figure*}

\subsubsection{Data augmentation (TCN-FTA, TCN-FSA)}

Data augmentation techniques are useful for artificially increasing the number of training data points, which can be of great benefit in cases of low or insufficient data such as ours. In order to evaluate the impact of data augmentation in our models, we adopted a simple strategy inspired by~\cite{Bock2020tcn, Pinto2021finetuning} in the experiments conducted with the TCN model: computing the input STFTs with different frame rates, i.e., varying hop sizes, so as to even out the distribution of tempi in the train set. Instead of randomly sampling from a normal distribution around the annotated tempo, we selected a set of frames rates $\pm2.5$\% and $\pm5$\% around its value. This allowed us to increase our sample size five-fold while maintaining the same amount of annotation effort. Models obtained with the data augmentation procedure are labeled TCN-FSA and TCN-FTA, for the training strategies described in Sections~\ref{sec:from_scratch} and~\ref{sec:fine_tuning}.

\subsection{Baselines}
\label{ssec:baselines}

We include two types of baselines. Firstly, the BayesBeat statistical model~\cite{Krebs2013bayesbeat} is used as reference to the adaptability and computational cost of the TCN. It has fewer parameters, thus training is faster. The second type of baselines are three off-the-shelf models---a Signal Processing technique, and two neural networks trained on Western music---; they illustrate the need for tailor-made solutions/adaptations in our context. Details %are 
presented below.

\noindent\textbf{BayesBeat.} This is based on the \emph{dynamic bar pointer model}~\cite{Whiteley2006}, and it simultaneously estimates beats, downbeats, tempo, meter, and rhythmic patterns, by expressing them as hidden variables in a hidden Markov model (HMM). An observation feature based on the spectral flux is computed from the audio signal and an observation model uses Gaussian mixture models (GMM) that are fitted during training to the feature values of each bin in a one--bar grid, so that rhythmic patterns are learned. Several patterns can be modeled, though one pattern is assumed to remain constant throughout the audio signal. 

BayesBeat has a few hyperparameters that the user should choose depending on the music. Those are the  number of rhythmic patterns, the type of feature to use (e.g., using only low, or low and high frequencies), and the feature grouping (e.g., how to compute the rhythmic pattern clusters), the tempo range, and whole note subdivisions. In \cite{Krebs2013bayesbeat}, it is reported that using two separate frequency bands ($\gtrless$ 250~Hz) %\qty{250}{\Hz}) 
helps finding the correct metrical level and is beneficial for beat and downbeat tracking. But, considering more frequency bands did not improve the results~\cite{Krebs2013bayesbeat}. According to~\cite{Holzapfel2014}, using one rhythmic pattern per rhythm class is usually enough to achieve a good performance and provides the best results in most cases. Following this, we use one rhythmic pattern and two frequency bands.

\noindent \textbf{Off-the-shelf baselines.}
We use the joint beat and downbeat tracking model of B\"ock et al. \cite{Bock2016} as per its implementation in \textit{madmom v0.17.dev0}~\cite{madmom}. It consists of an LSTM-based model trained in ten datasets spanning Western genres, and Carnatic, Cretan and Turkish music excerpts. We also include the beat tracker from Ellis \cite{Ellis2007}, which estimates a global tempo and then uses dynamic programming to find the best set of beats that reflect such tempo. As a final baseline, we include the TCN of Section \ref{ssec:tcn} trained with the Western datasets from Section \ref{ssec:datasets} (TCN-BL).

\subsection{Evaluation metrics}

We use as our main metric F-measure \cite{Dixon2007}, along with the continuity-based metrics \cite{hainsworth2003techniques, Klapuri2006} CMLt (“correct metrical level”) which corresponds to the ratio between correct and annotated beats, and AMLt (“allowed metrical level”) which accepts phase errors of half a beat period or octave errors in estimation. For the computational cost of the models, we simply report the time they take to train by using in-build timing functions in the code. 

\section{Experiments and Results}

\begin{table*}[ht]\small
    \centering
    \begin{tabular}{l|ccc|ccc|ccc|ccc}
       \hline
        & \multicolumn{6}{c|}{\emph{Candombe}} & \multicolumn{6}{c}{\emph{Samba}} \Tstrut \\
       \cline{2-13}
        Model & \multicolumn{3}{c|}{Beat} & \multicolumn{3}{c|}{Downbeat}
        & \multicolumn{3}{c|}{Beat} & \multicolumn{3}{c}{Downbeat} \Tstrut\\
        \cline{2-13}
         & CMLt & AMLt & F & CMLt & AMLt & F &
        CMLt & AMLt & F & CMLt & AMLt & F \Tstrut \\\hline
        BayesBeat\_0.67     & 95.0 & 95.0 & 96.1 & 82.2 & 92.0 & 81.7 & 70.5 & 74.2 & 78.8 & 57.4 & 72.9 & 57.9 \Tstrut \\
        BayesBeat\_12.33    & 99.6 & 99.6 & 99.6 & 99.8 & 99.8 & 99.4 & 93.5 & 96.0 & 96.7 & 92.5 & 94.9 & 92.9 \\
        BayesBeat\_all      & 98.6 & 98.6 & 98.9 & 98.8 & 98.8 & 98.4 & 94.0 & 96.0 & 96.9 & 92.0 & 95.3 & 92.5 
        \\\hline
        TCN-FSA\_0.67       & 88.2 & 89.1 & 90.2 & 56.5 & 70.4 & 61.6 & 74.4 & 78.2 & 79.9 & 24.3 & 60.1 & 41.8 \Tstrut \\
        TCN-FSA\_12.33      & 97.9 & 98.0 & 98.4 & 95.8 & 98.0 & 95.4 & 94.3 & 96.3 & 97.0 & 92.8 & 97.5 & 95.7 \\
        TCN-FSA\_all        & 98.4 & 98.4 & 98.6 & 97.4 & 98.4 & 97.5 & 96.0 & 98.6 & 98.2 & 95.0 & 96.6 & 95.9 \\\hline
        TCN-FTA\_0.67       & 96.7 & 96.7 & 97.3 & 81.8 & 89.7 & 82.1 & 85.3 & 93.0 & 91.7 & 28.5 & 81.5 & 60.9 \Tstrut \\
        TCN-FTA\_12.33      & 99.4 & 99.4 & 99.4 & 96.5 & 99.5 & 96.1 & 95.8 & 97.8 & 98.1 & 92.0 & 97.2 & 95.0  \\
        TCN-FTA\_all        & 99.4 & 99.4 & 99.4 & 97.4 & 99.5 & 97.1 & 96.3 & 98.4 & 98.2 & 96.1 & 97.7 & 96.0 \\\hline
        TCN-BL              & 11.1 & 18.7 & 15.9 & 14.9 & 31.9 & 4.1 & 46.5 & 65.6 & 60.0 & 5.9 & 52.5 & 9.6 \Tstrut  \\
        Ellis~\cite{Ellis2007}   & 34.8 & 38.1 & 38.0 &    - &    - &   - & 82.3 & 87.6 & 87.1 &   - &    - & - \\
        B\"{o}ck \cite{Bock2016} & 11.7 & 14.4 & 11.5 & 26.7 & 40.3 & 0.5 & 46.9 & 76.0 & 66.4 & 5.2 & 66.6 & 2.0 \\
        \hline
    \end{tabular}
    \caption{Mean F-measure (F) and continuity scores (CMLt, AMLt) in beat and downbeat tracking tasks across both genres.}
    \label{tab:evaluation_results}\vspace*{-.5\baselineskip}
\end{table*}

\subsection{Performance of models}

Figure \ref{fig:fmeasure} shows the F-measure results for the TCN models trained for \emph{candombe} and for \emph{samba} with different amounts of data using each of the training strategies, as well as BayesBeat, computed as the bootstrapped results of ten experiments (95\% confidence) with different random seeds for each combination of model and data amount.

A first striking observation is that for both beats and downbeats, the performance curve for most models has a small positive slope, which means it is indeed possible to nearly achieve best model performance (which would require training with full dataset) by just training with few samples. This is particularly true for the estimation of beat, for which models rapidly reach F-measure scores above 80\% with %less than a minute
1.5~min %\qty{1.5}{\minute}
of data in both \textit{candombe} and \textit{samba} for %almost
all configurations.
This is an interesting result, meaning that not much gain in performance is expected with the increase of annotations for such datasets. An end-user could annotate %less than 
about a minute of data and yet obtain decent performances. The same holds for downbeat in the best performing models for \emph{candombe}, but not in \emph{samba}. For %In
the latter, there is a clear gain in adding more data, which has to do with the differences between the two rhythms, as discussed below.

\noindent \textbf{Differences between \emph{candombe} and \emph{samba}.} 
Observing the results in Figure \ref{fig:fmeasure}, we see that the models tend to require more data to achieve better performance on \emph{samba} than on \emph{candombe}. %, and the uncertainty about the performance for \emph{samba} is larger. 
Our intuition behind this result is that, as mentioned in Section \ref{subsec:afro_music}, because \emph{samba} has a bigger combination of timbres and pitches than candombe, the decision of what snippets to annotate (i.e., the sampling) might be more critical for the former than for the latter, e.g., ensuring timbre representation. 

\noindent \textbf{Best model configuration.} 
The best performing configuration for beat and downbeat tracking in both music traditions is the fine-tuned TCN model with data augmentation (FTA). Particularly, data augmentation produced significant improvement in performance for downbeat tracking in samba. Interestingly, for the adaptive setting concerned in this work, the BayesBeat baseline is competitive with the TCN model, especially considering the computational cost (see Section~\ref{sec:comp_cost}).

\noindent \textbf{Comparison with off-the-shelf benchmarks.} 
Table~\ref{tab:evaluation_results} shows the performance of the TCN and the BayesBeat baseline for different data subsets, namely the smallest and largest subsets, and the full dataset. It also shows the performance of the three off-the-shelf baselines explained in Section~\ref{ssec:baselines}. In alignment with previous works~\cite{Nunes2015candombe, Maia2018brid}, the models trained with Western music (TCN-BL and B\"ock~\cite{Bock2016}) perform very poorly in \emph{candombe}, and reach only about 66\% F-measure in \emph{samba}, both significantly lower than the performance of the same models in Western music genres. The model of Ellis~\cite{Ellis2007} scores considerably better, but is not consistent in both datasets. This shows the necessity of adapting meter tracking models to these music genres, as even the models trained with the smallest subsets of data (0.67 and 1.5~min) %\qty{1.5}{\minute})
outperform the baselines.

\subsection{How much time do the models take to train?}
\label{sec:comp_cost}

\begin{figure}[ht]
    \centering
    \includegraphics[scale=.47]{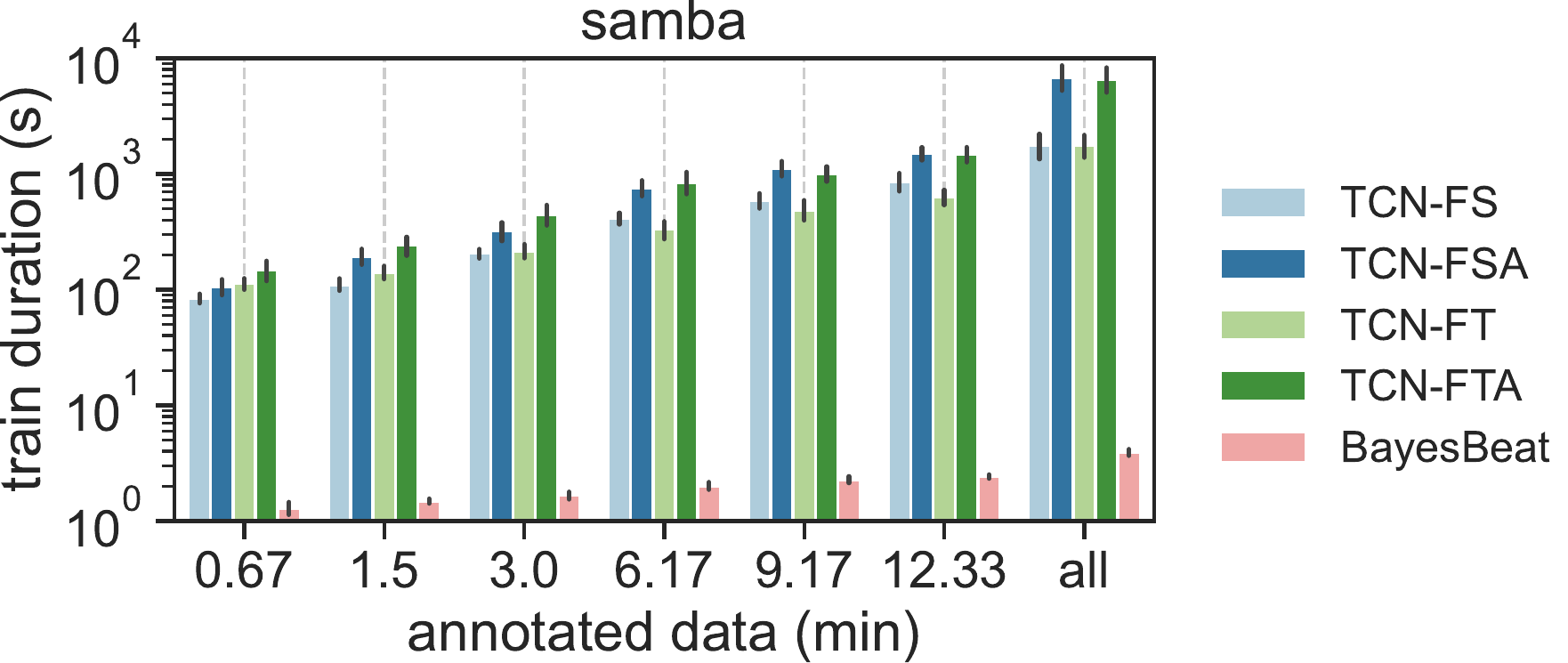} %.52
    \caption{Training time for the different amounts of data. }
    \label{fig:comp_cost}\vspace*{-.5\baselineskip}
\end{figure}

Our analysis is motivated by the adaptation of meter tracking models in real-world use cases. For this adaptation to make sense it has to be done quickly. In this regard, we estimate the time each model configuration takes in training, and contrast it with the BayesBeat baseline. Figure~\ref{fig:comp_cost} shows how the train duration varies with the size of the train set for \emph{samba} (very similar results were obtained for \emph{candombe}). The TCN takes about the same time in both \emph{samba} and \emph{candombe}, with a minimum of about 100~s %\qty{100}{\second}
for the smallest subset. Among the TCN configurations, the most expensive ones use data augmentation. This makes sense given that more data is used for training. As expected, the BayesBeat trains significantly faster than the TCN, taking on average 1.62~s %%\qty{1.62}{\second}
to train with 0.67~min %\qty{0.67}{\minute}
of data, and being in the order of 50 to 350 times faster than the TCN when data augmentation is not used. This big gap in computing time, together with the results of Figure~\ref{fig:fmeasure} and Table~\ref{tab:evaluation_results}, makes BayesBeat an overall good alternative for adapting meter tracking to these Latin American music. We observed that all configurations take about the same inference time, around 20~s %\qty{20}{\second}
for the full test set.

\subsection{When can we train with small data?}

\begin{figure}
    \centering
    \begin{subfigure}{\linewidth}
    \centering
    \includegraphics[scale=.47]{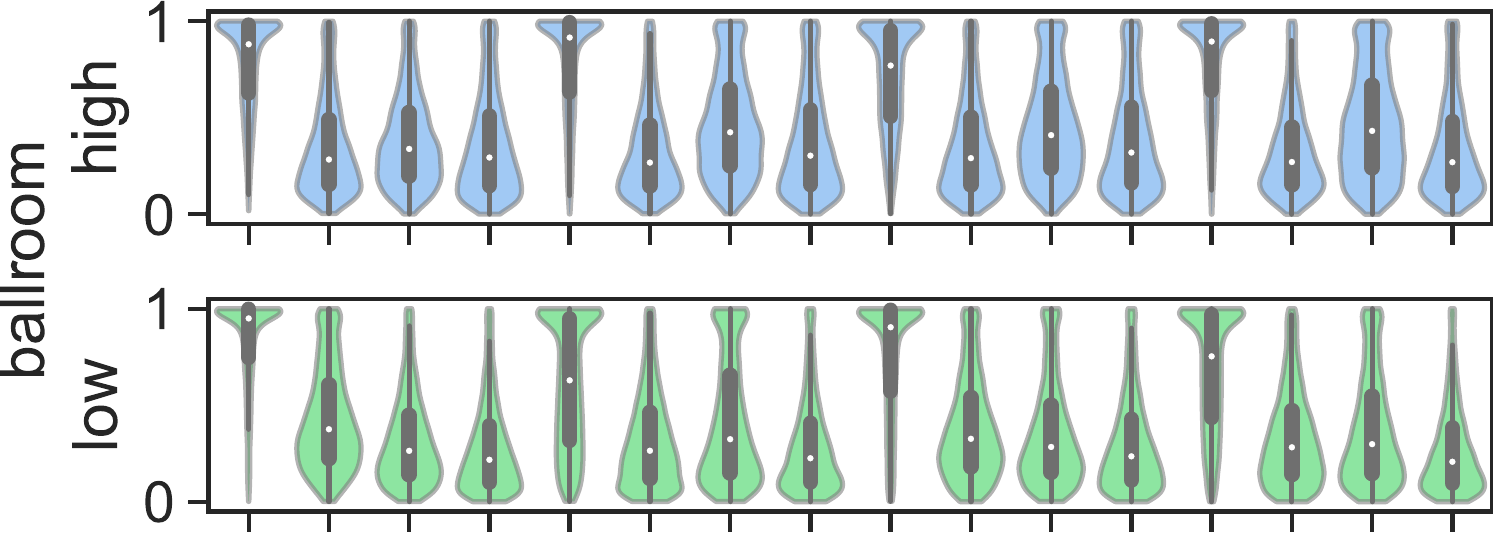}
    \label{fig:violin_downbeats_ballroom}
    \end{subfigure}\vspace*{5pt}
        \begin{subfigure}{\linewidth}
    \centering
    \includegraphics[scale=.47]{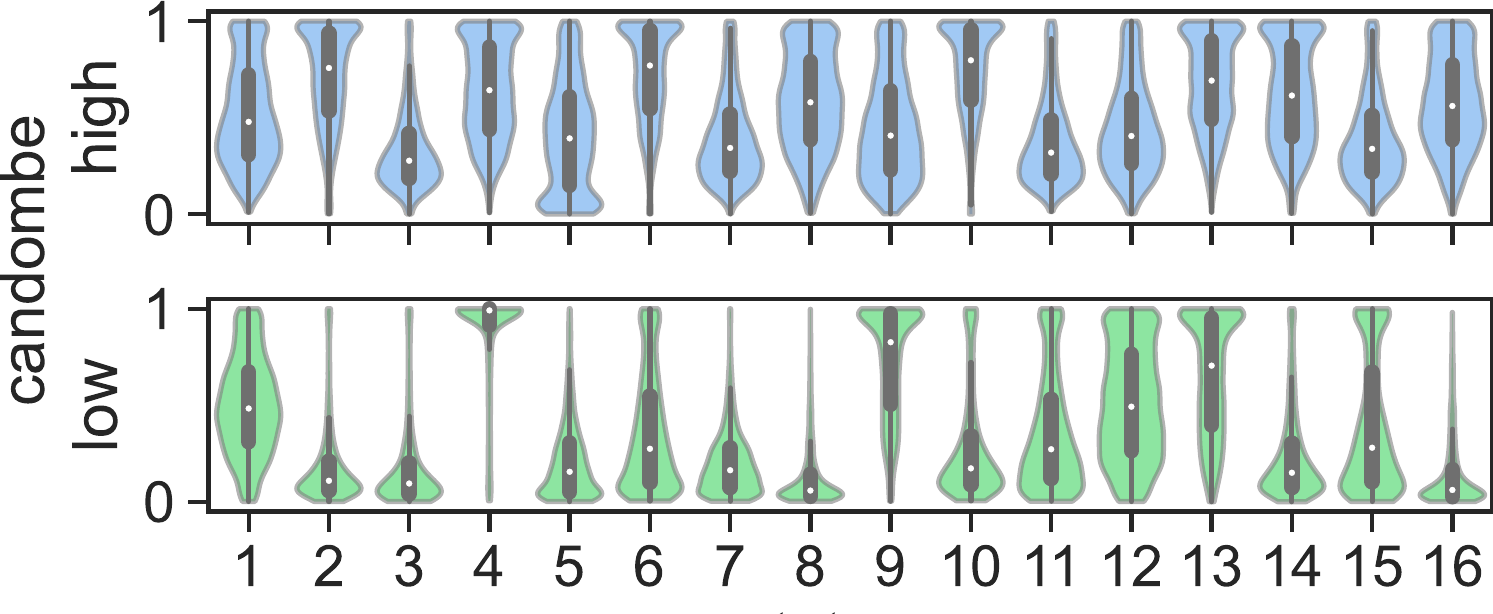}
    \label{fig:violin_downbeats_candombe}
    \end{subfigure}\vspace*{-2pt}
    \begin{subfigure}{\linewidth}
    \centering
    \includegraphics[scale=.47]{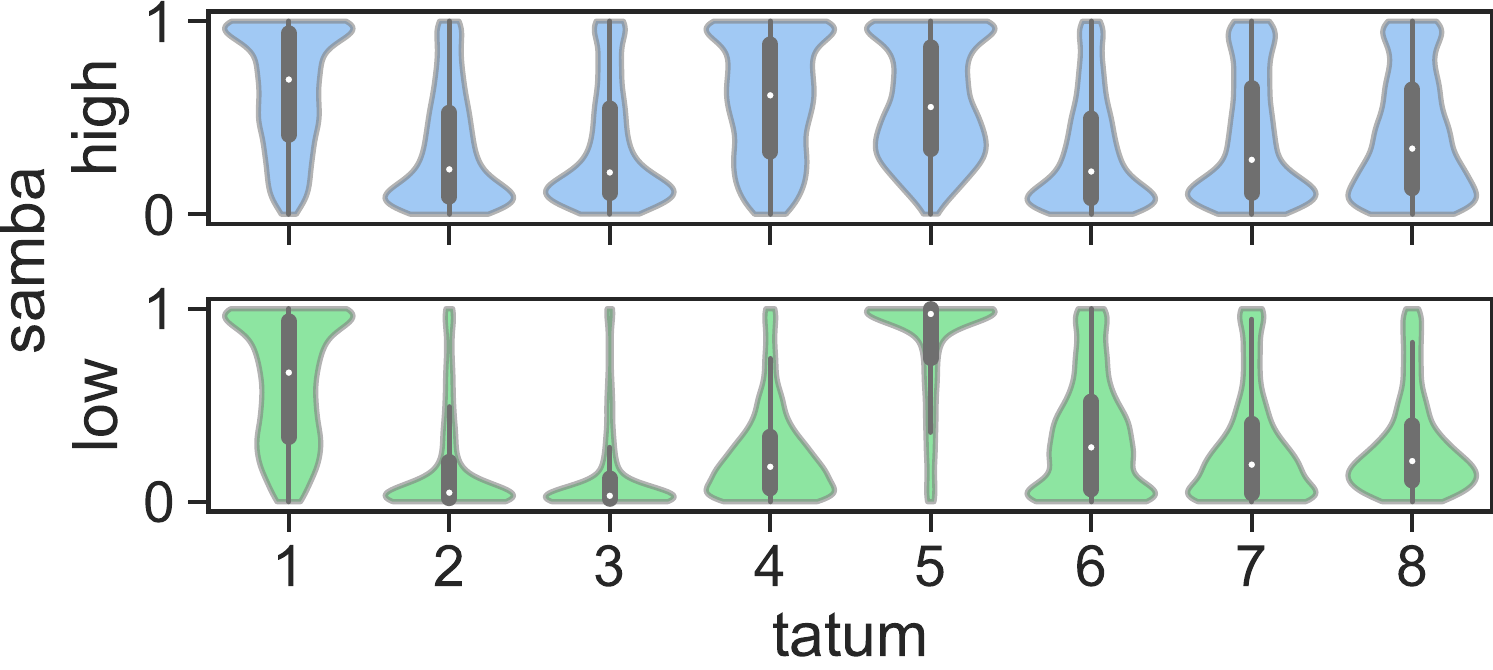}
    \label{fig:violin_downbeats_samba}
    \vspace*{-6pt}
    \end{subfigure}
    \caption{Tatum strength distribution per frequency band for Ballroom (just 4/4 tracks), \emph{candombe}, and \emph{samba}.}
    \label{fig:violin}\vspace*{-.5\baselineskip}
\end{figure}

Our intuition is that the more variability in the data (in terms of meters, rhythmic patterns and instrumentation), the harder it is for a model to learn with small data. This aligns with our experiments in the adaptability of these methods to \emph{samba} and \emph{candombe}, and also agrees with the musicological insights of Section~\ref{subsec:afro_music}. To have a more quantitative understanding of this, we derived a bar profile for each type of music. First we extract a feature map from each excerpt using the beat/downbeat annotations to time-quantize a locally normalized onset strength function~\cite{Rocamora2014} at the tatum scale --- this was done with the \emph{carat}~\cite{carat} toolbox, considering the tatum duration as one quarter of the time-span between successive beats. % as the fourth part of a beat
Then, for each dataset, we summarize these feature maps across time, which results in a distribution of feature values per tatum. To allow an analysis of these profiles in different regions of the spectrum, we compute the onset strength in two frequency bands (20~Hz to 200~Hz; %\qtyrange{20}{200}{\hertz};
and > 200~Hz).%\qty{200}{\hertz}).
We present these distributions as violin plots in Figure~\ref{fig:violin} for \emph{candombe}, \emph{samba}, and for the Ballroom dataset.

In Figure 3, we verify that for some tatums strength distributions are concentrated around 1 or 0, indicating a strong characteristic accent or lack thereof at that point of the bar respectively. High variance, in its turn, means ``fuzzyness’’ in the rhythm pattern, which could justify the difficulty in learning that rhythm, specially with small data.

\emph{Samba}, which has eight tatums per bar (2/4 meter), is known for having a strong metrical accent at beat 2, which we may readily identify in its low-frequency channel at tatum 5. The first beat also has a high median value but is less ``deterministic’’ due to its high variance. In turn, the low-frequency profile of \emph{candombe} displays a high-variance downbeat, no accent on beat 2, and strong accents on beats 3 and 4, but also a strong contrametric accent at tatum 4. These characteristics could help explain why the off-the-shelf beat tracking models, which expect beats to be accented, perform worst on \emph{candombe}. Looking back at \emph{samba}, we see that tatums 2 and 3 show small standard deviations and correspond to ``off’’ tatums; together with beat 2, they make three out of eight tatums that exhibit very small variance in the low channel. In \emph{candombe}, besides tatum 4, tatums 2, 3, 7, 8, 9, 14, and 16 also present small variance. This abundance of ``anchor’’ points could justify why adaptation in \emph{candombe} came with little data.

\begin{figure}[!ht]
    \centering
    \includegraphics[scale=.47]{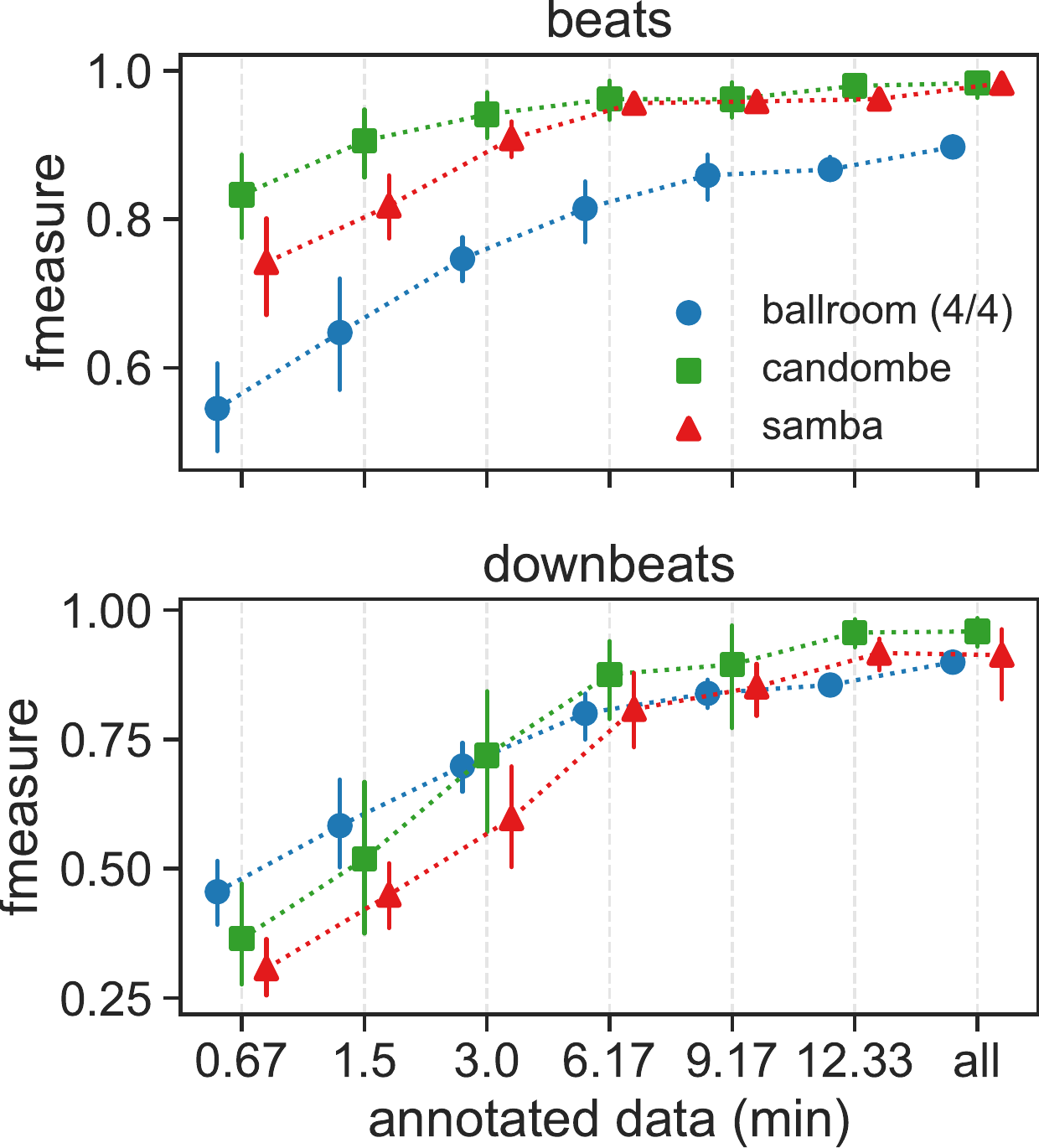}
    \vspace*{-6pt}
    %\caption{TCN performance in Ballroom with small data.} % added after meta
    \caption{TCN-FS performance in Ballroom.}
    \label{fig:ballroom}\vspace*{-.5\baselineskip}
\end{figure}

In Ballroom, we clearly see that beats are distinct for having high strength and low variance in both channels, whereas the rest of the tatums show no clear trend. Its few reference points could pose a challenge for learning models. Furthermore, beat patterns (the combination the four tatums in-between beats, including the beat itself) are also indistinguishable from one another, which could aggravate this matter. To test these observations, we trained a set of models from scratch for Ballroom using the same methodology that for \textit{samba} and \textit{candombe}. Results are depicted in Figure \ref{fig:ballroom}. The performance results correlate with the intuition that Ballroom is a more challenging dataset, particularly for beat tracking, given that it comprises multiple genres, and also that for learning beat and downbeat more data would be needed.

\section{CONCLUSIONS AND FUTURE WORK}

We adapted a meter tracking model using small quantities of data to work in particular Latin American music traditions, namely \textit{samba} and \textit{candombe}. We showed that, under certain homogeneity conditions, it is indeed possible to train such models with a few minutes of annotated data and training cycles, and obtain almost full-dataset performance. This result has promising consequences in real-world applications, as it opens the possibility of adapting such models to other music genres with modest labeling efforts. The most competitive model is a fine-tuned TCN with data augmentation, whereas BayesBeat is a good option under computational cost constraints. In the future, we will investigate rhythm complexity metrics that could serve to predict the amount of annotated data needed to adapt meter tracking models to particular music genres.

\section{ACKNOWLEDGEMENTS}

This study was partially funded by the Coordenação de Aperfeiçoamento de Pessoal de Nível Superior - Brasil (CAPES) --- Finance Code 001. % and by the Agencia Nacional de Investigación e Innovación - Uruguay (ANII).

\bibliography{ISMIRtemplate}

\end{document}